# Quantum Uncertainty and Nonlocality: Are they Correctly Understood?

Elias P. Gyftopoulos

In a brief article [1], Seife refers to works by Einstein and Schrödinger and concludes that there is a relentless murmur of confusion underneath the chorus of praise for quantum theory. It is noteworthy that a "murmur" is not necessarily a cause for replacement of any scientific theory, and that the issues raised by Einstein, Podolsky, and Rosen, and Schrödinger's responses to the EPR paper have been satisfactorily resolved by Gyftopoulos and von Spakovsky [2] in a manner that renders the relentless murmur mute and unwarranted. In what follows, I respond briefly to specific issues raised in [1] where it is asserted that:

(i) "…quantum theory is bizarre and counterintuitive…. The equations of quantum mechanics work very well, they just don't seem to make sense". Such characterizations are not consistent with the nature of physical reality so eloquently defined by Margenau [3]. If intuition were a criterion of validity of any scientific theory then our solar system would still be geocentric, and the observations of Aristarhus (circa 300 BC) and Galileo (circa AD 1500) would continue being counterintuitive.

(ii) "…an atom, for example, can be on the left side of a box and the right side of a box as long as the atom is undisturbed and unobserved. But as soon as an observer opens the box and tries to spot where the atom is, the superposition collapses and the atom instantly chooses whether to be on the right or on the left". Such conclusions are not part of quantum theory. They are based on von Neumann's projection postulate which is shown not to be valid [4, 5], and the misconception that a probability distribution can be disregarded as a result of a single measurement. No probability distribution either of statistics or of the unified quantum theory of mechanics and thermodynamics [2] can be determined by one measurement.

(iii) "…the cat paradox is used by Schrödinger to ridicule superposition". But superposition is not a quantum mechanical concept either at one instant in time or, even more so, at two instants in time [2]. A ket (density operator) can be represented by an *expansion* in terms of a complete set of orthonormal kets (orthonormal projectors). For either of the two expansions, there exists an infinite number of complete sets.

(iv) "…in the classical world an object has a solid reality, whereas in quantum theory the solid reality is undermined". The Uncertainty Principle expressed as "if one gains knowledge about position, one loses knowledge about momentum" is used as a justification for the assertion just cited. However, this is a completely unjustified conclusion [2]. The uncertainty principle requires two sets of measurements performed on two ensembles of identical systems, identically prepared, one for the probability distribution of position and the standard deviation of this distribution, and the second for similar results about momentum. The two standard deviations are then used in the uncertainty relation that corresponds to the noncommuting operators for position and momentum. So, no loss of knowledge is involved in these two measurements.

(v) "…two particles fly away from each other and wind up at opposite ends of the galaxy. But the two particles happen to be entangled – linked in a quantum-mechanical sense – so that one particle instantly feels what happens to its twin. Measure one, and the other is instantly transformed by that measurement as well; it's as if the twins mystically communicate, instantly, over vast regions of space. This nonlocality is a mathematical consequence of quantum theory and has been measured in the lab. The spooky action apparently ignores distance and the flow of time". We discuss in great detail these assertions [2], and prove that they are false because they violate the foundations of quantum theory.

(vi) "The mathematical framework is sound and describes all these bizarre phenomena well. If we humans can't imagine a physical reality that corresponds to our equations, so what? That attitude has been called the shut up and calculate interpretation of quantum mechanics". I find this assertion nonscientific. All rational theories of physics are judged on the basis of whether they can regularize our perceptions by experimentally verifiable predictions. If they fail either in logic or in experimental verifiability, or both then we must start looking for a better paradigm. In either case, bizarreness is not a criterion.

1. Massachusetts Institute of Technology, 77 Massachusetts Ave., Cambridge, MA 02139